\documentclass{elsart}

\usepackage{epsfig}
\usepackage{natbib}

\journal{Advances in Space Research}

\begin{document}

\begin{frontmatter}

\title{A statistical study of hot flow anomalies using Cluster data}
\author[rmki]{G. Facsk\'{o}\corauthref{cor}}, 
\corauth[cor]{Corresponding author.}
\ead{gfacsko@rmki.kfki.hu}
\author[rmki]{K. Kecskem\'{e}ty}, 
\author[rmki]{G. Erd\H{o}s}, 
\author[rmki]{M. T\'{a}trallyay}, 
\author[lindau]{P. W. Daly},
\author[toulouse]{I. Dandouras}

\address[rmki]{KFKI Research Institute for Particle and Nuclear Physics, H-1525 Budapest, PO.Box 49, Hungary}
\address[lindau]{Max Planck Institute for Solar System Research, Max-Planck-Str. 2, 37191 Lindau-Katlenburg, Germany}
\address[toulouse]{CESR, 9, Avenue du Colonel ROCHE, 31028 Toulouse CEDEX 4, France}

\begin{abstract}
Hot flow anomalies (HFAs) are studied using observations of the RAPID suprathermal charged particle detector, the FGM magnetometer, and the CIS plasma detector aboard the four Cluster spacecraft. Previously we studied several specific features of tangential discontinuities on the basis of Cluster measurements in February-April 2003 and estimated the size of the region affected by the HFA in different ways. In this paper we confirm the following results: the angle of Sun direction and tangent ional discontinuity (TD) normal must be larger then $45^o$, the magnetic field directional change is small and their estimated size is between $2.20\pm0.28\,R_{Earth}$. We then present evidence for a new necessary condition for the formation of HFAs, that is, the \textit{solar wind speed is significantly (about 200\,km/s or $\Delta M_f=2.3$) higher} than the average. The existence of this condition is also confirmed by simultaneous ACE MAG and SWEPAM solar wind observations at the L1 point 1.4 million km upstream of the Earth. The results are confronted with recent hybrid simulations.
\end{abstract}


\end{frontmatter}

\section{Introduction}
\label{sec:intro}

Hot flow anomalies still seem to be one of the mysteries of magnetospheric physics since their discovery \citep{schwartz85, thomsen86:_hot}. Two different ways exist to come to know an unknown phenomenon: the first and so charming way of performing case studies about single events, and the more difficult and longer one to analyze more events and making general assumptions on them. We chose the last one because \citet{schwartz00:_condit}'s previous global investigation and our previous studies also indicated that HFAs are not special and rare phenomena \citep{kecskemety06:_distr_rapid_clust}. On the contrary, they frequently appear when the necessary conditions occur, however, before Cluster no spacecraft could prove this because the statistics was insufficient. ESA's new fleet completely changed this situation allowing us and other research groups to identity more events and examine them properly \citep{lucek04:_clust}. 

\begin{figure}[t]
\centering
\epsfig{file=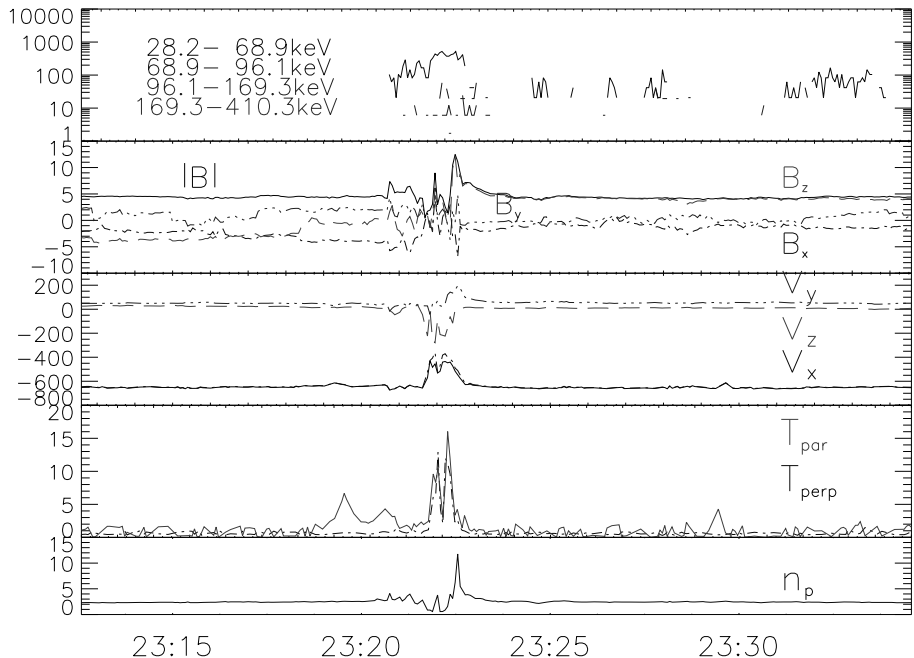, width=350pt}
\caption{\textit{A hot flow anomaly event occurred on 23:22 (UT) March 23, 2003 as measured by Cluster-1 RAPID, FGM and CIS instruments. First panel on the top: energetic particle fluxes (in $p/{cm}^{3}\,s\,sr\,keV$). Second panel: magnetic field components (in nT). Third panel: solar wind speed components  (in GSE, km/s). Fourth panel: solar wind temperature parallel and perpendicular to $\underline{B}$ (MK). Bottom panel: plasma density ($cm^{-3}$).}}
\label{fig:hfa}
\end{figure}

It is commonly agreed that HFAs are generated by the interaction of a tangential discontinuity and the bow-shock. They are explosive events, particle acceleration, magnetic depletion take place at their center and the magnitude of the magnetic field increases at the rims. The plasma temperature increases whereas the density drops within them, but the most interesting phenomenon is the directional change of the solar wind flow (\textit{Fig.~\ref{fig:hfa}}). The flow turns away from the anti-sunward direction and might even be directed backward \citep{thomsen86:_hot, thomsen93:_obser_test_hot_flow_abnom, sibeck99:_compr, sibeck02:_wind}.

Only a few theoretical approaches have been devoted to the description of HFAs \citep{burgess89, thomas91:_hybrid, lin97:_gener, lin02:_global}. The newest one is a hybrid simulation by \citet{lin02:_global}. It is very difficult to confront its results directly to the observations. We chose a clear geometrical prediction to check: the size of HFA depends on the angle between the TD normal and Sun-Earth direction ($\gamma$). Lin predicts a monotonous increase up to $80^{o}$ and then a fall off. The size of the HFA depends on angular change of the magnetic field direction at the TD ($\Delta\Phi$). Prediction: a monotonically increasing function. 

The structure of the paper is the following: in Section~\ref{sec:dep} we discus our investigation about the HFA size dependence on TD parameters, in Section~\ref{sec:fast} we report a new HFA formation condition and finally in Section~\ref{sec:summary} we gives the summary of our results.

\section{The dependence of HFA size on TD properties}
\label{sec:dep}

We set a short series of criteria based on \citet{thomsen86:_hot, thomsen93:_obser_test_hot_flow_abnom, sibeck99:_compr, sibeck02:_wind} that were:
\begin{enumerate}
\item The rims of the cavity must be visible as a suddenly increase of magnetic field magnitude compared to the unperturbed solar wind region's value. Then the magnetic filed value drops and its direction vector turns around. At the rim of the cavity it increases again and reaches its original value. 
\item The solar wind speed drops, sometimes it may flow back to the Sun direction but its direction always turns away from the Sun-Earth direction. 
\item The solar wind temperature increases and its value reaches several ten million degrees.
\item The solar wind particle density also increases on the rims of the cavity and drops inside HFA.
\item Energized particle flux can often - but not always - be observable at HFA. The increased flux starts before the magnetic signatures and ends after them. 
\item Finally the calculated convective electric field must point to the TD on both side of the discontinuity. 
\end{enumerate}
See also \textit{Fig.~\ref{fig:hfa}}. In this way we found more then 50 HFA events \citep{kecskemety06:_distr_rapid_clust}. Unfortunately we could not determine many discontinuity normal using Cluster FGM and ACE MAG measurements so finally we worked only 33 events in our study. 

\begin{figure}[t]
\centering
\epsfig{file=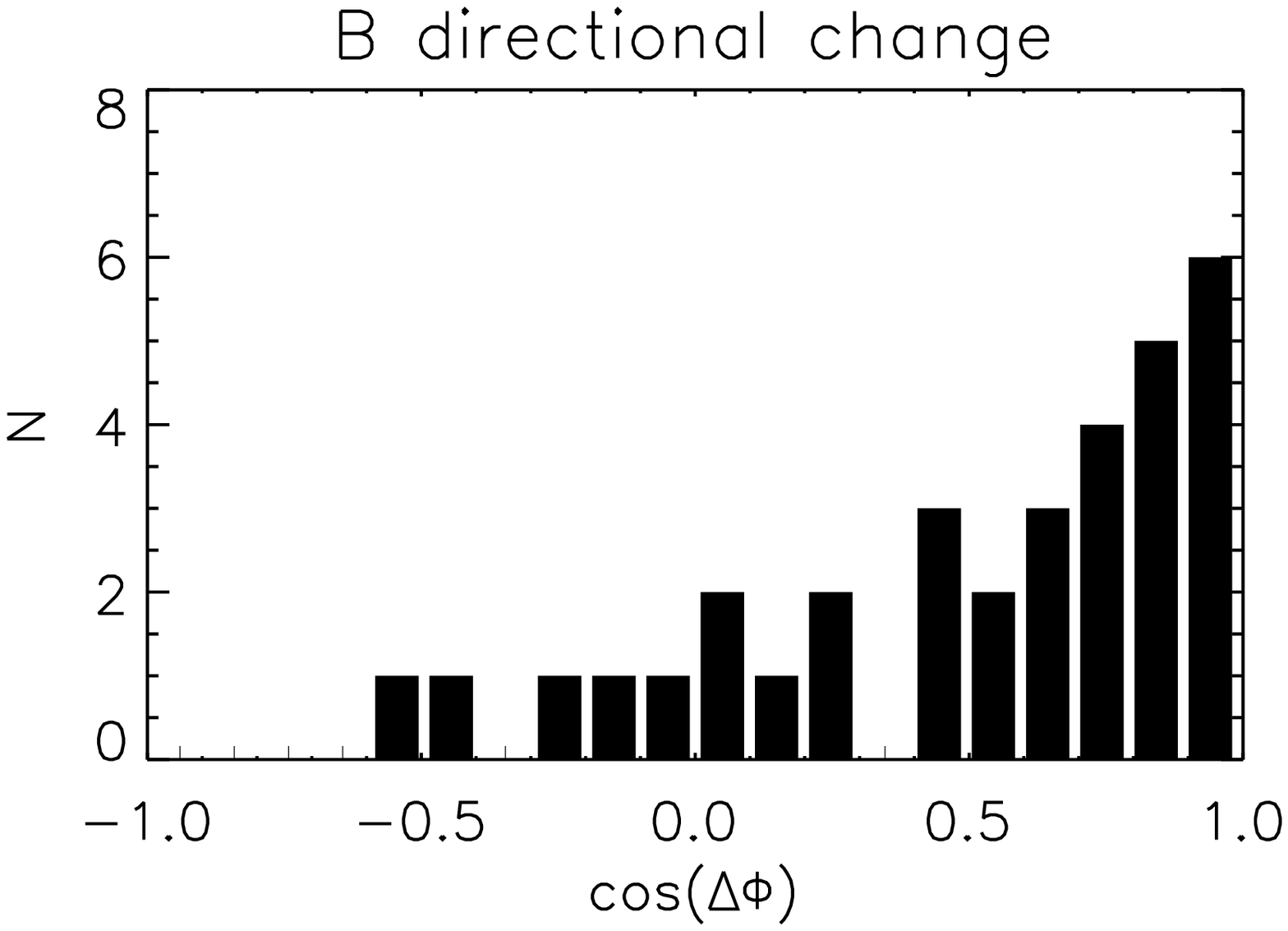, width=175pt}
\epsfig{file=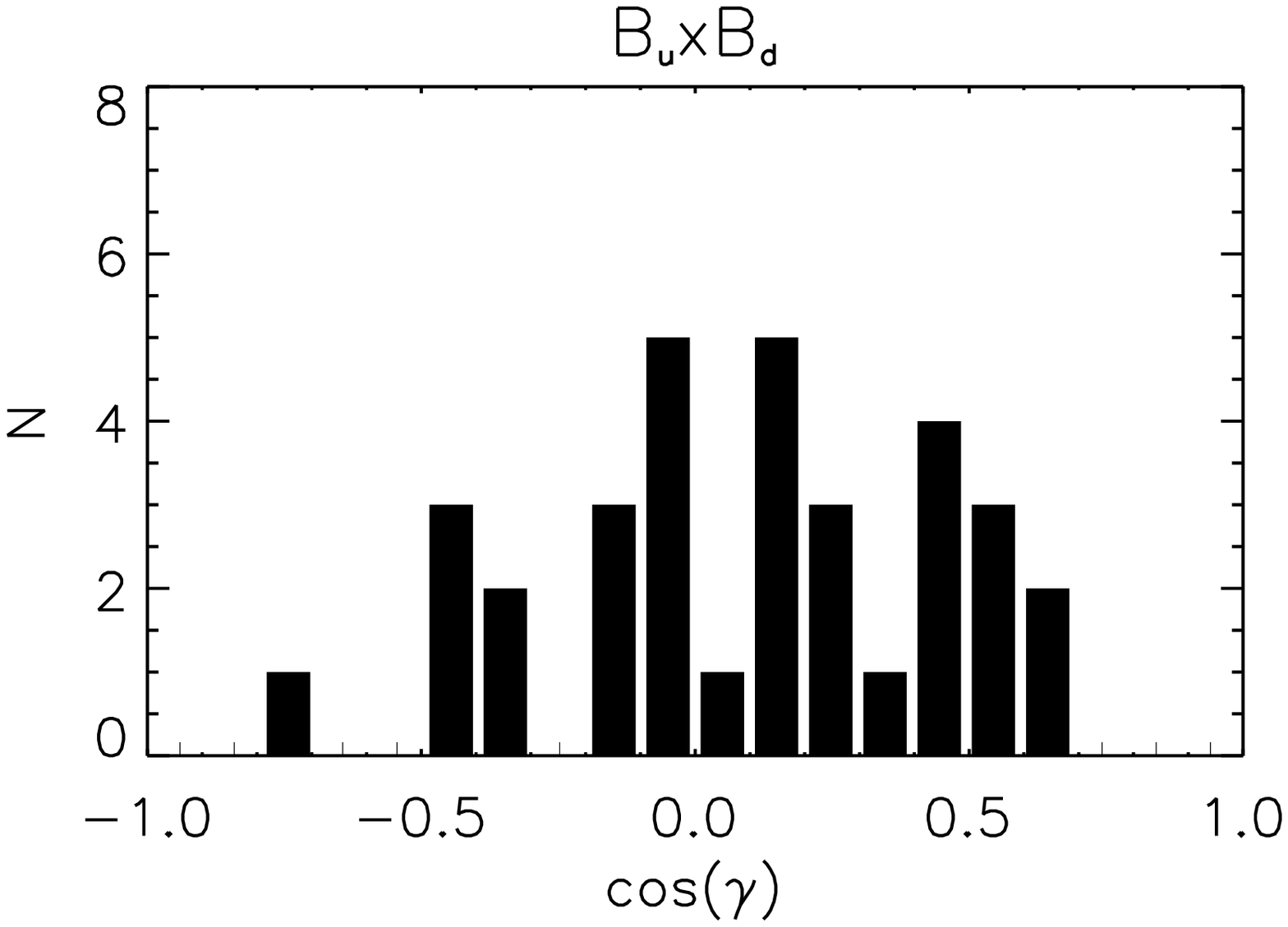, width=175pt}
\caption{\textit{Distribution of $\cos \left(\Delta\Phi\right)$ (left) and $\cos \left(\gamma\right)$ (right) where $\gamma$ is the angle of TD normal vector and the Sun direction; furthermore $\Delta\Phi$ is the angle of magnetic field direction change on both sides of the discontinuity. We used $\cos$ instead of pure angles to get less twisted distributions.}}
\label{fig:directions1}
\end{figure}

\begin{figure}[t]
\centering
\epsfig{file=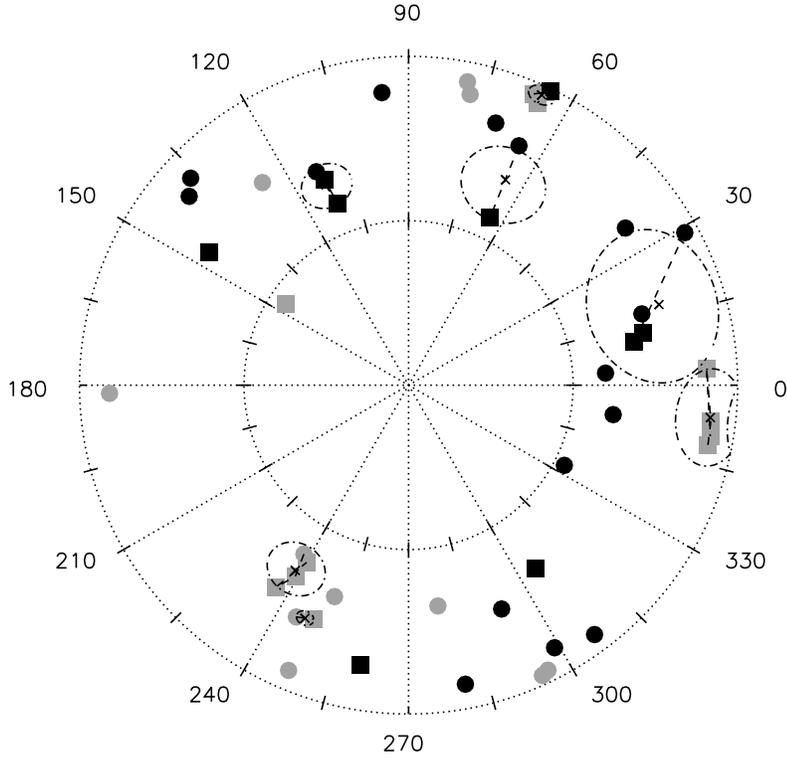, width=300pt}
\caption{\textit{Polar plot of the direction of the normal vectors of the TDs. The azimuthal angle is measured between the GSE y direction and the projection of the normal vector onto the GSE yz plane. The distance from the center is the cone angle of the normal vector of the TD determined by the cross-product method. The regions surrounded by dashed lines are the projection of error cones. Circles and squares symbolize ACE and Cluster data, respectively.}}
\label{fig:directions2}
\end{figure}

We determined the tangential discontinuities which generated HFA events between February 1 and April 16, 2003 using Cluster-1 and -3 FGM 1s average magnetic field and CIS 4s average solar wind speed measurements as well as MAG magnetic (16s) and SWEPAM (64s) plasma data from ACE. We applied both the minimum variance techniques \citep{sonnerup98:_minim_maxim_varian_analy} and the cross product method to determine their normal vectors. We determined whether the normal component of the magnetic field dissolves at the discontinuity. If its value approaches zero then the discontinuity was tangential and we could use the cross product method to determine its normal vector. The minimum variance method is usually quite inacurate so we accepted its result if angle of the minimum variance and the crossproduct vector was less then $15^o$; furthermore the rate of the 2nd and 3rd was greater than $2.0$. We calculated the angles for which theoretical predictions were available and confronted the simulation results with observations. The $\Delta\Phi$ values seemed to be large (${}^o\pm{}^o $) inside the TD (\textit{Fig.~\ref{fig:directions1}, left}) and  the existence of a wide gap in the cone angle (about $45^{o}$) around the Sun-Earth direction was confirmed (\textit{Fig.~\ref{fig:directions1}, right}) \citep{lin02:_global}. The \textit{Fig.~\ref{fig:directions2}}\ emphasizes and visualizes better this large predicted and observed gap around the Sun direction. We plotted all discontinuity normal determined using Cluster FGM and ACE MAG measurements. The Cluster measurements are symbolized by squares and the circles symbolize normal vectors calculated using ACE data. If HFA was observed and its TD normal was calculated more than one Cluster and ACE spacecraft the average vector was plotted and the larger difference to its direction was considered as error and it was drawn by dot dashed line; furthermore this angle was also plotted on \textit{Fig.~\ref{fig:directions2}}. Finally normal belong to the same events were blinded by dashed line. 

\begin{figure}[th]
\centering
\begin{tabular}[t]{rl}
\epsfig{file=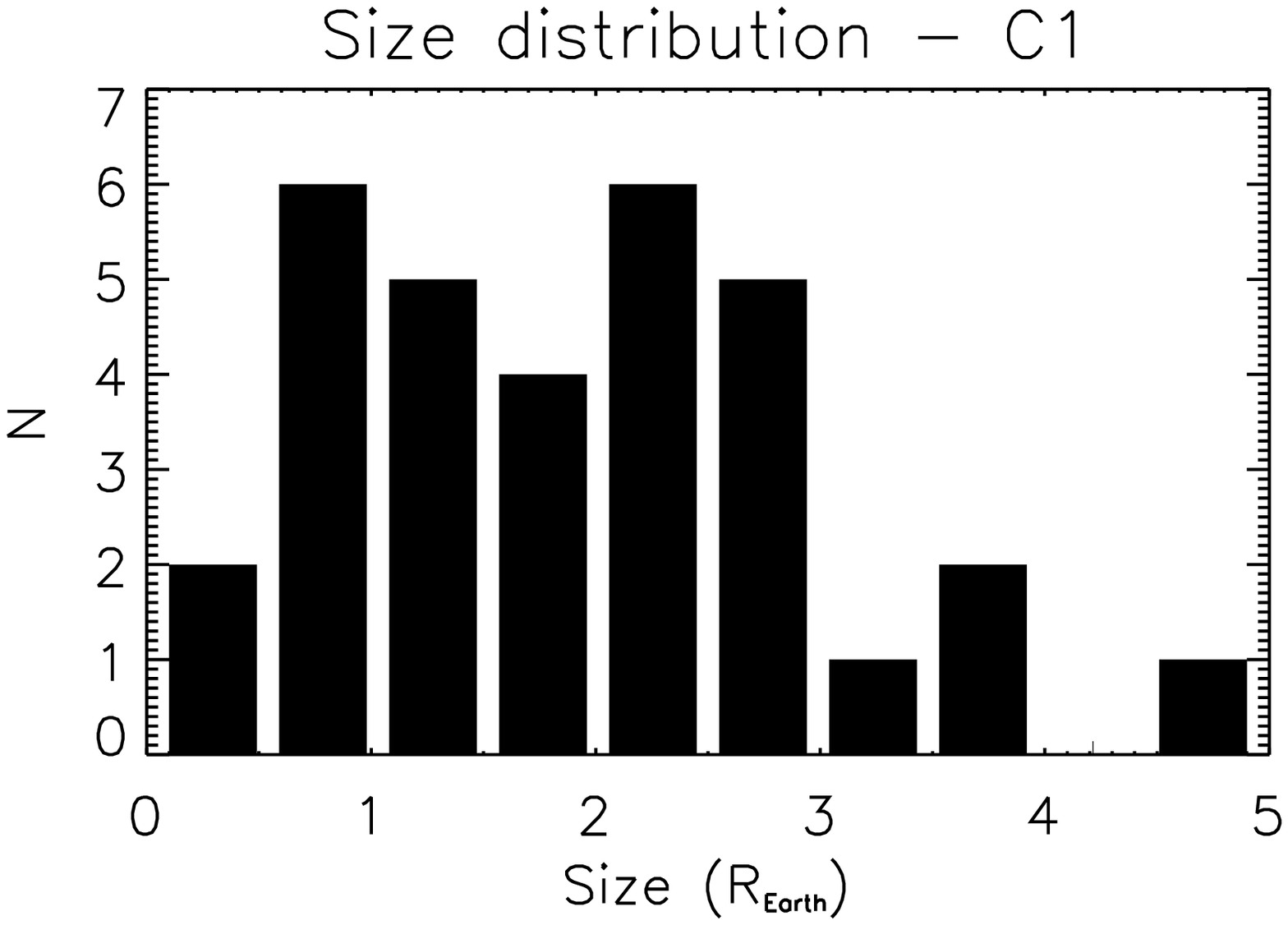, width=175pt} &
\epsfig{file=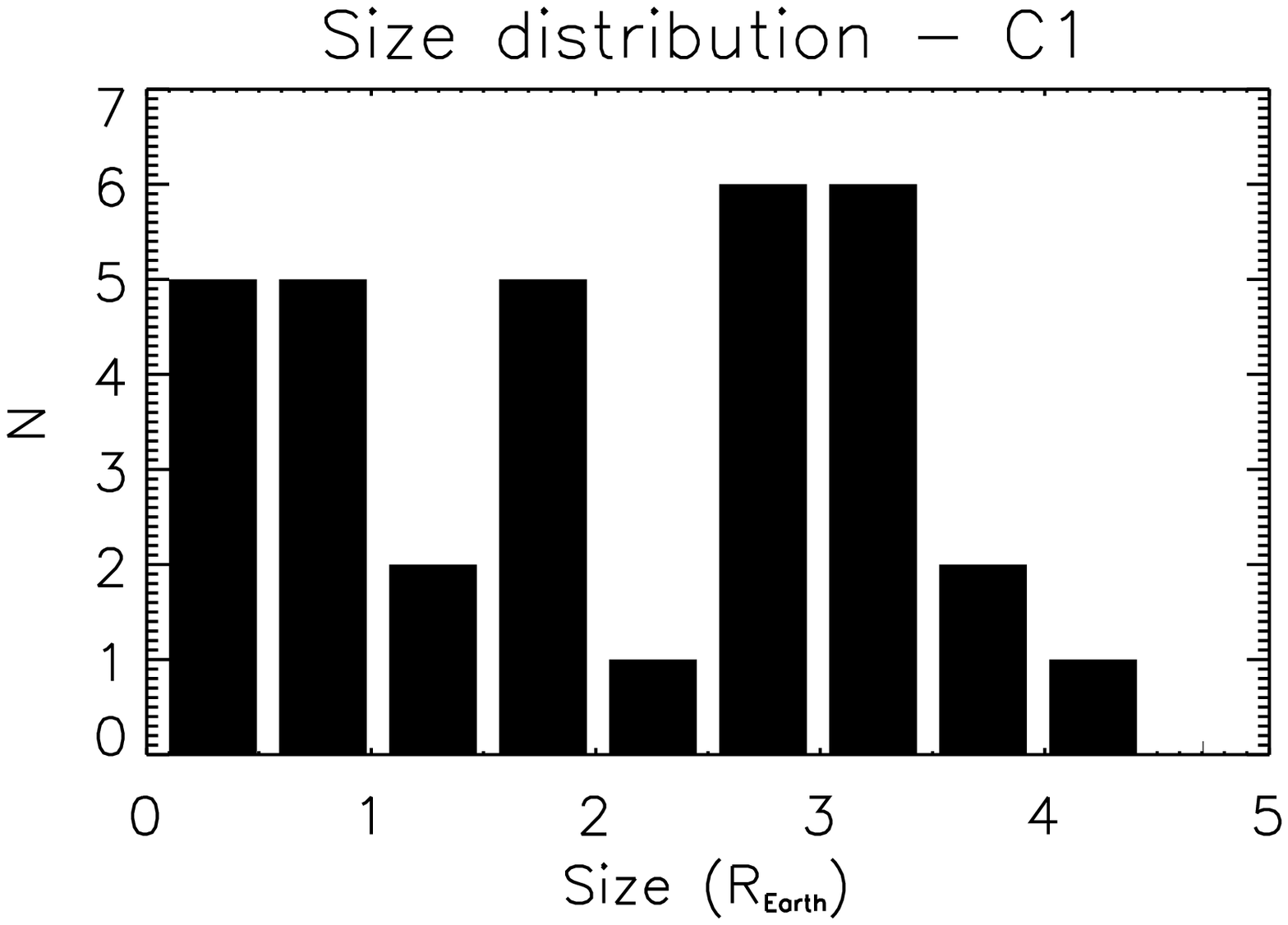, width=175pt} \\
\epsfig{file=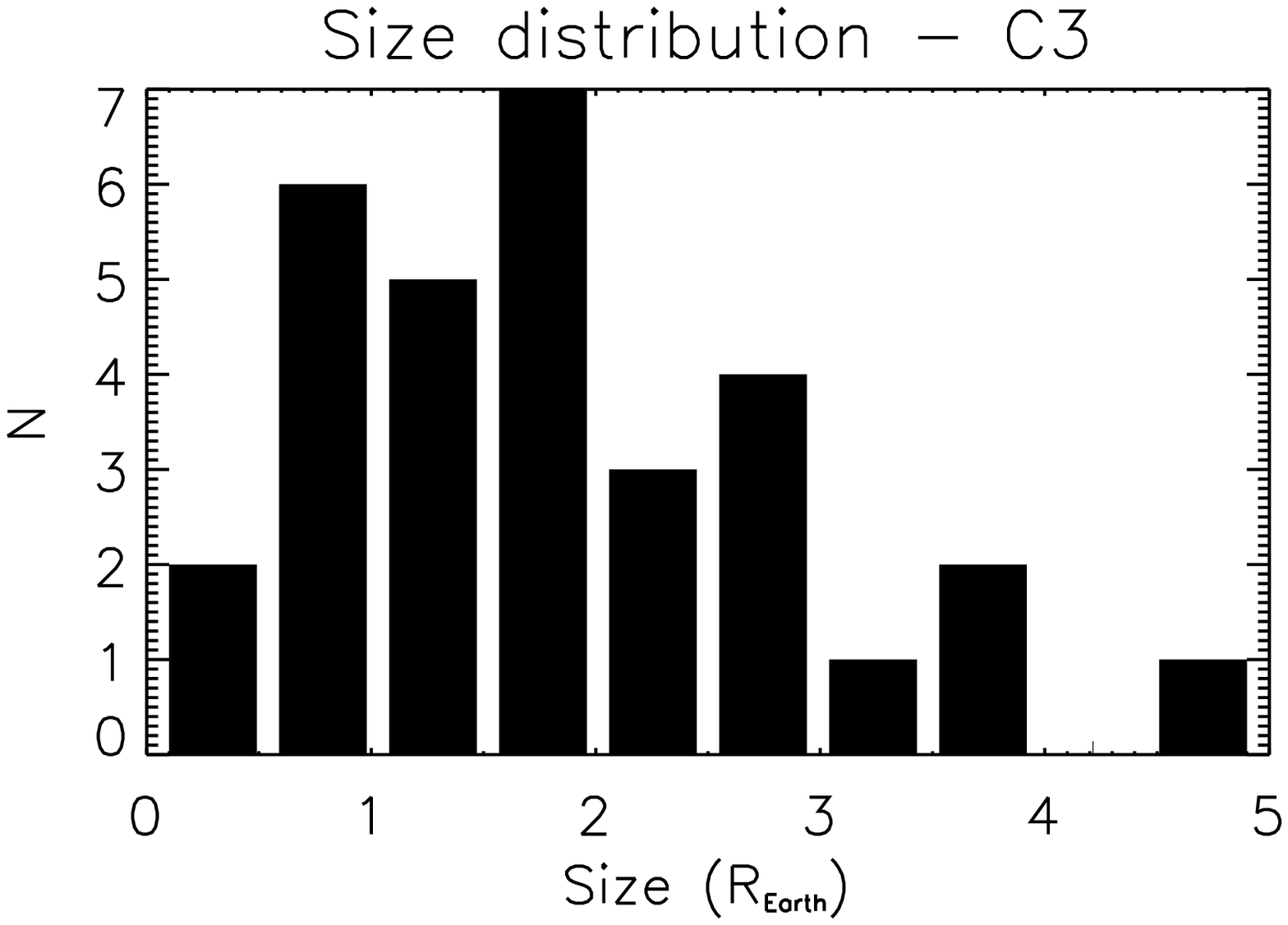, width=175pt} &
\epsfig{file=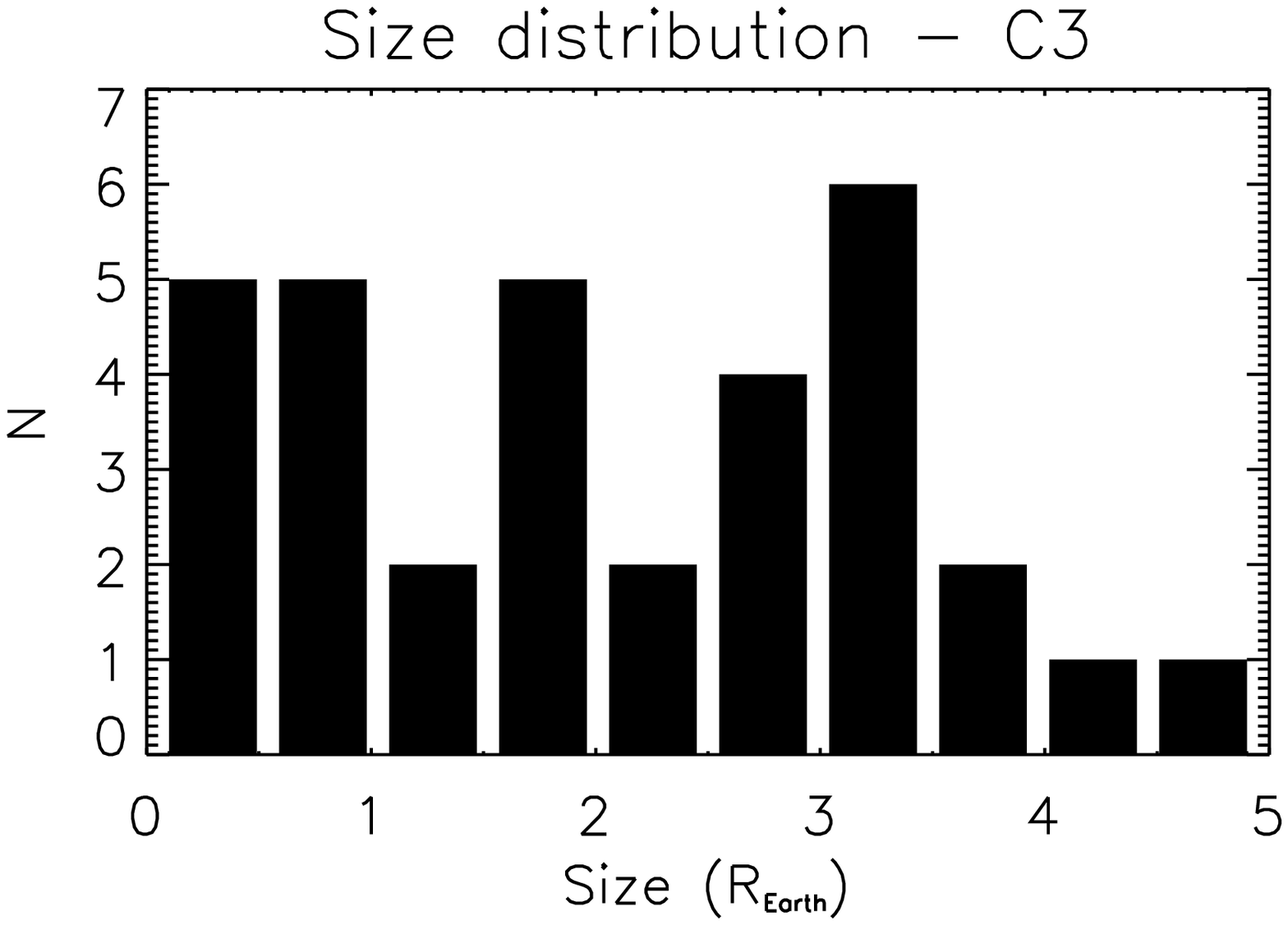, width=175pt} \\
\end{tabular}
\caption{\textit{Size distributions of hot flow anomalies determined using measurements of FGM and CIS aboard Cluster-1 and -3. Left panel: Distributions based on crossing time of HFA region and right: the separation of spacecraft.}}
\label{fig:size}
\end{figure}

We estimated the size of the region affected based on the separation and crossing time of the spacecraft in solar wind reference frame \citep{facsko05:_ident_statis_analy_hot_flow}. The principle of our methods was very simple and rough. When spacecraft cross the cavity the durations of intersection and the length of the orbit are determinated. Using CIS HIA solar wind speed measurements aboard Cluster-1 and Cluster-3 we transformed the intersection positions to the solar wind frame. The average length was considered as the diameter of HFA and the error was calculated based on their difference to the average.  The separation of the fleet helps us to perform the minimal size of the HFA event. We also transformed the intersection point positions to solar wind frame here however it was not so important in this case. The size and its error were calculated in the same way like in the case of crossing. In all case the error of size estimation is very sensitive for the configuration of the fleet and the shape of the cavity. The separation method does not give good result if the separation is small (in 2003 the separation was large). The result ($2.20 \pm 0.28 R_{Earth}$) which is the average of the four different results is in accordance with \citeauthor{lin03:_global}'s theoretical predictions (\textit{Fig.~\ref{fig:size}}). We gave this average to compensate the error of the different estimations given by the principle of methods. 

Actually, the accuracy of these estimation methods resulted in large errors and this fact did not allow us to agree or disprove \citeauthor{lin03:_global}'s geometrical  predictions. On the other hand, we discovered that the solar wind speeds were significantly higher than the average value when HFAs were generated. This suggests that the fast solar wind speed is an important factor of HFA formation.

\section{Fast solar wind as a favorable condition for HFA formation}
\label{sec:fast}

\begin{figure}[th]
\centering
\begin{tabular}[t]{rl}
\epsfig{file=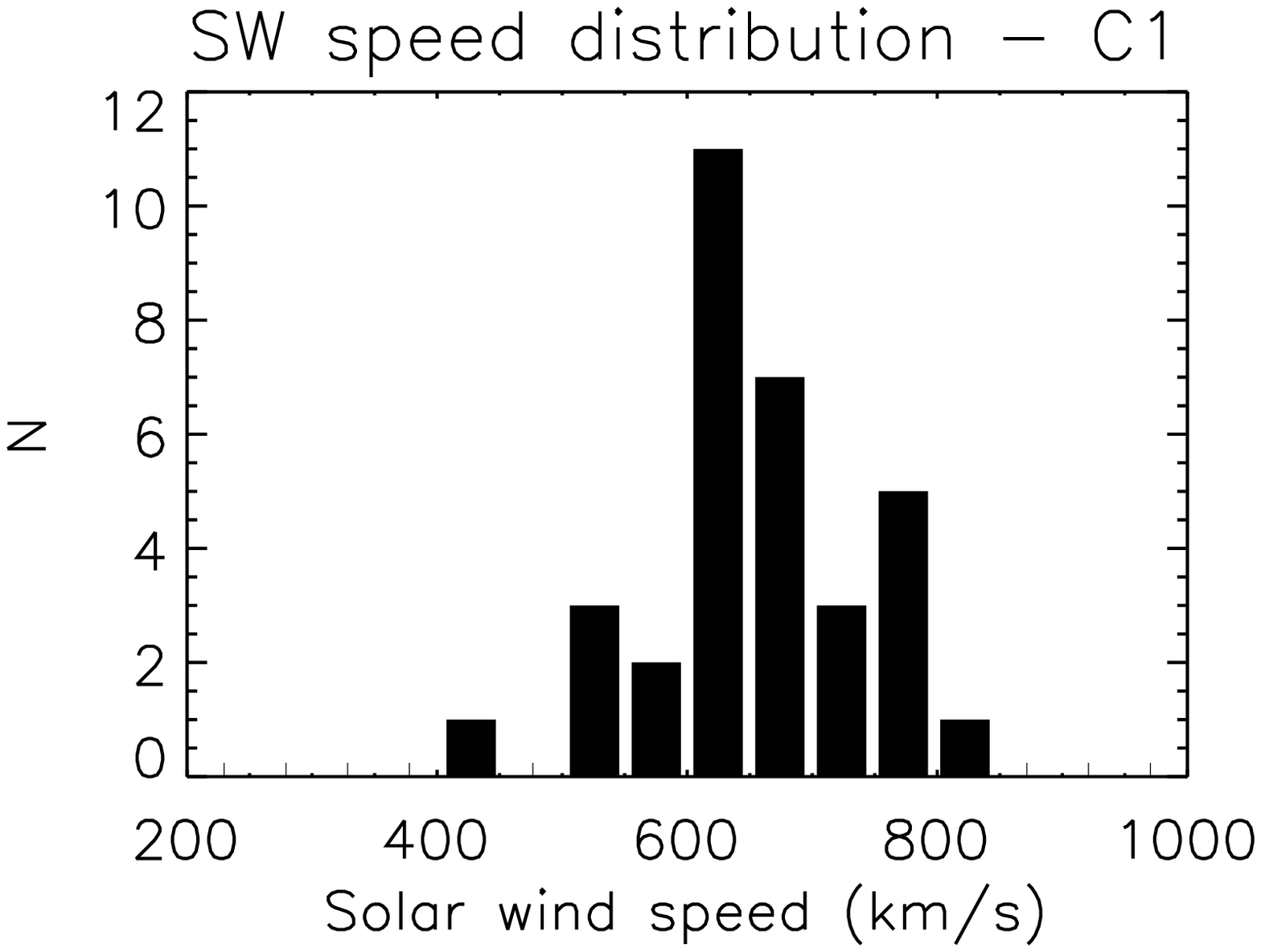, width=150pt} &
\epsfig{file=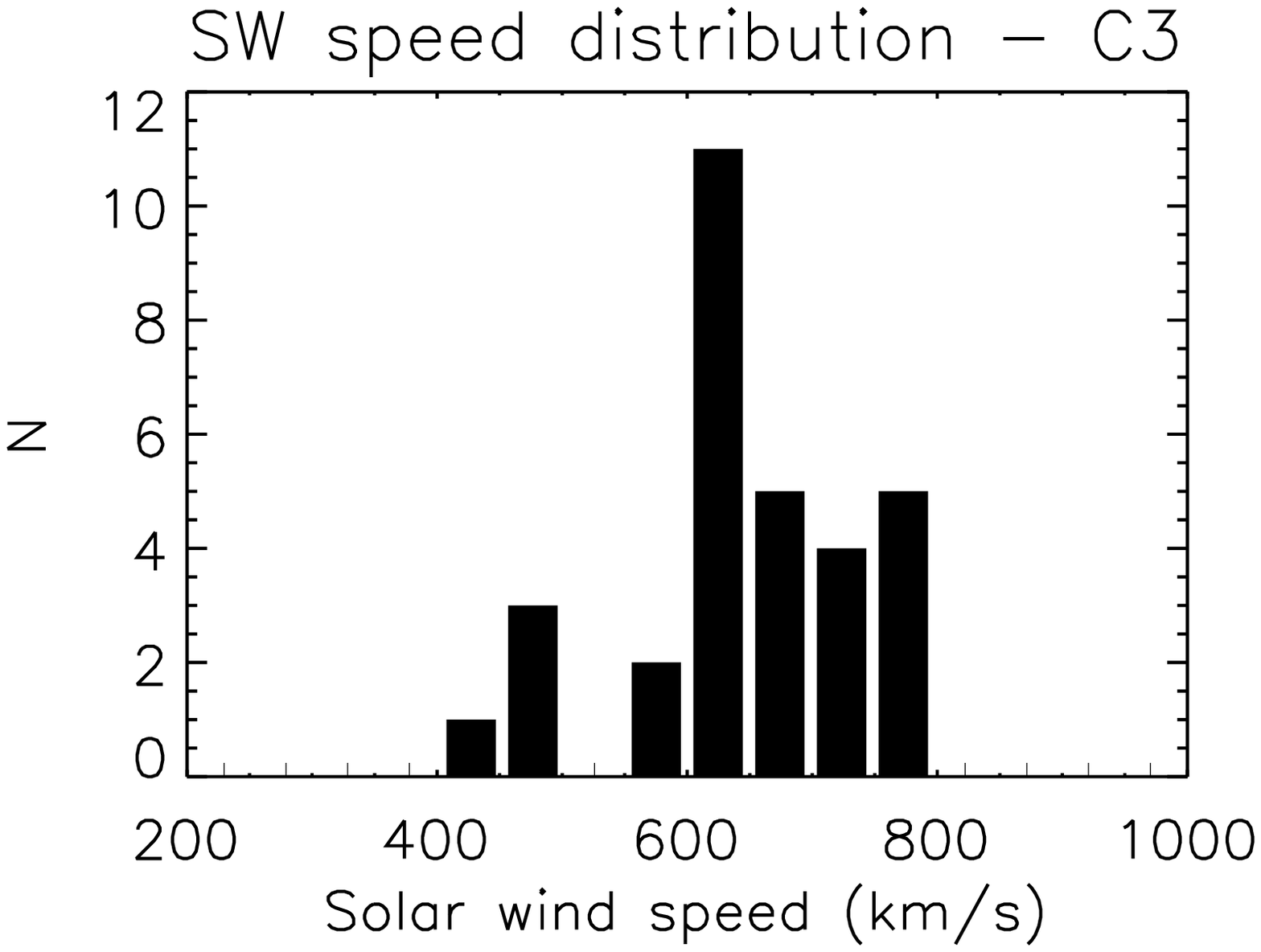, width=150pt} \\
\epsfig{file=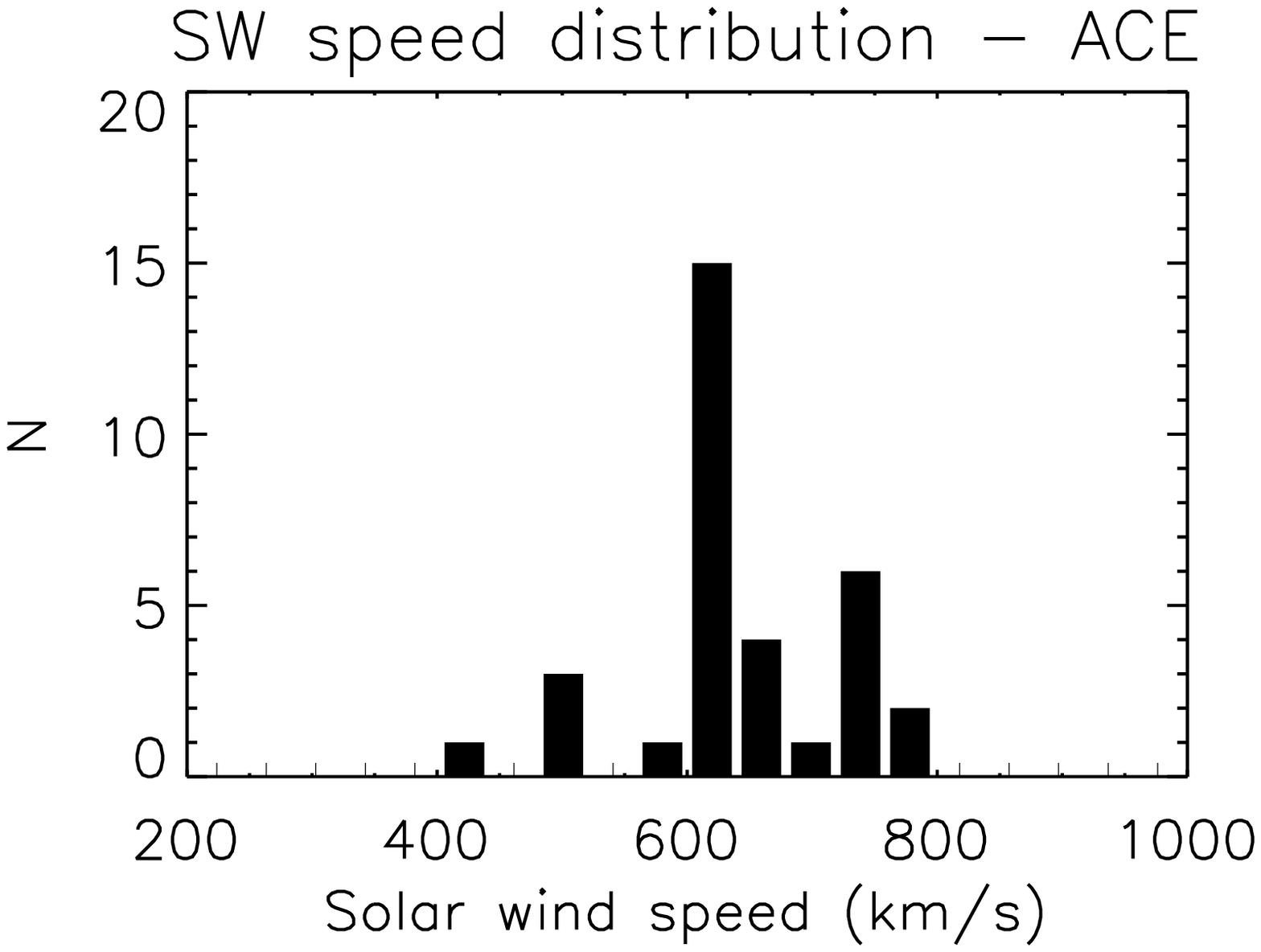, width=150pt} & 
\epsfig{file=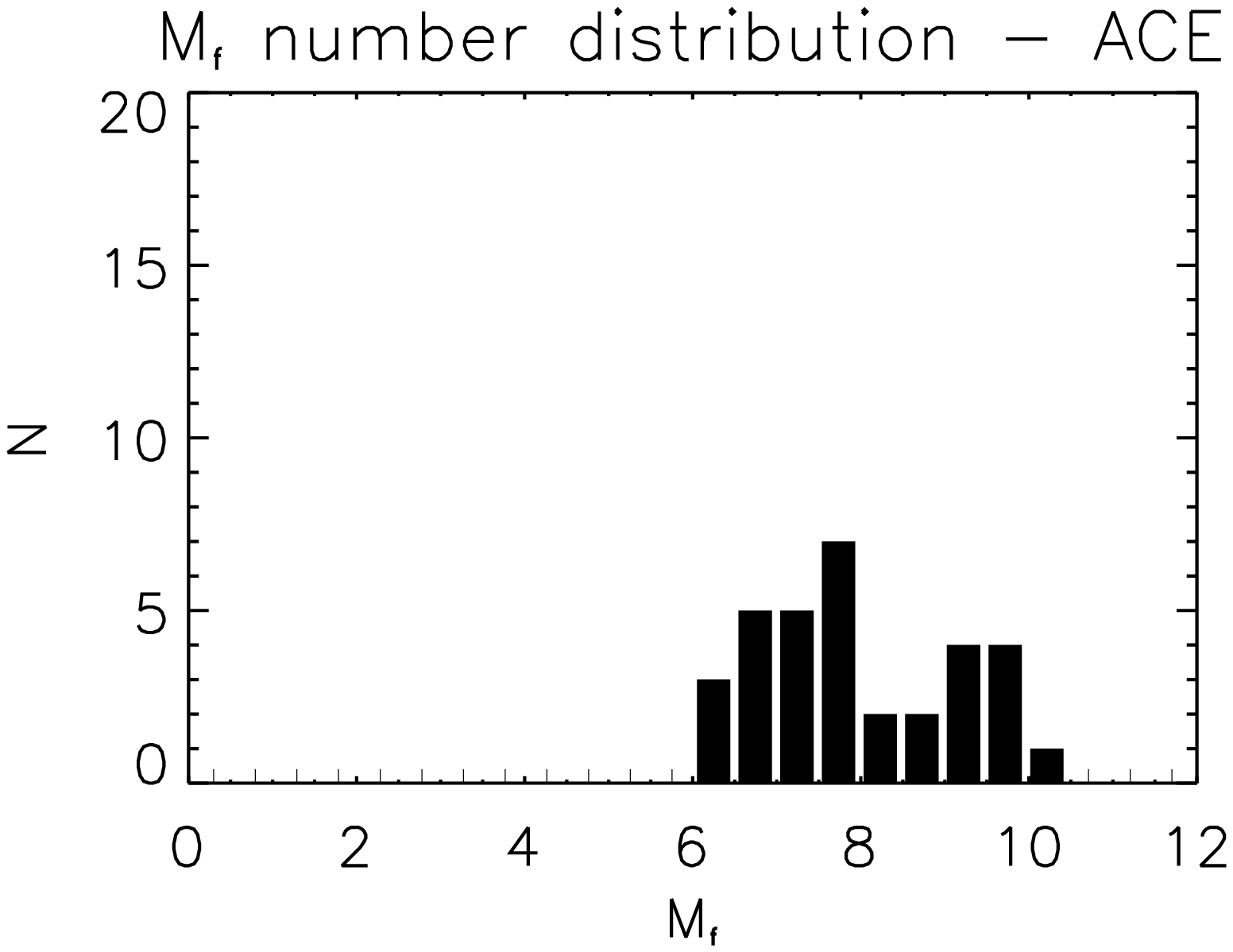, width=150pt} \\
\epsfig{file=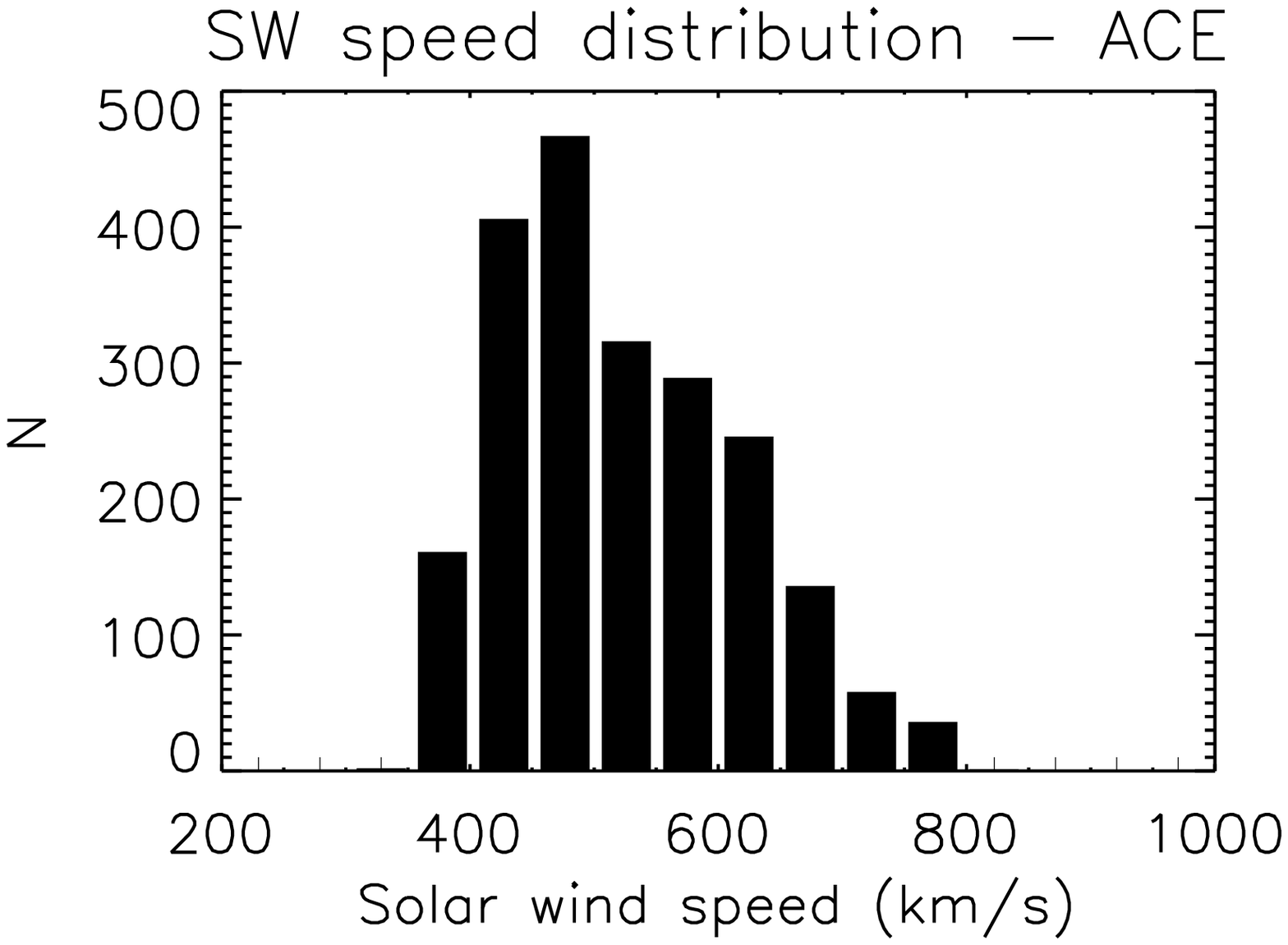, width=150pt} &
\epsfig{file=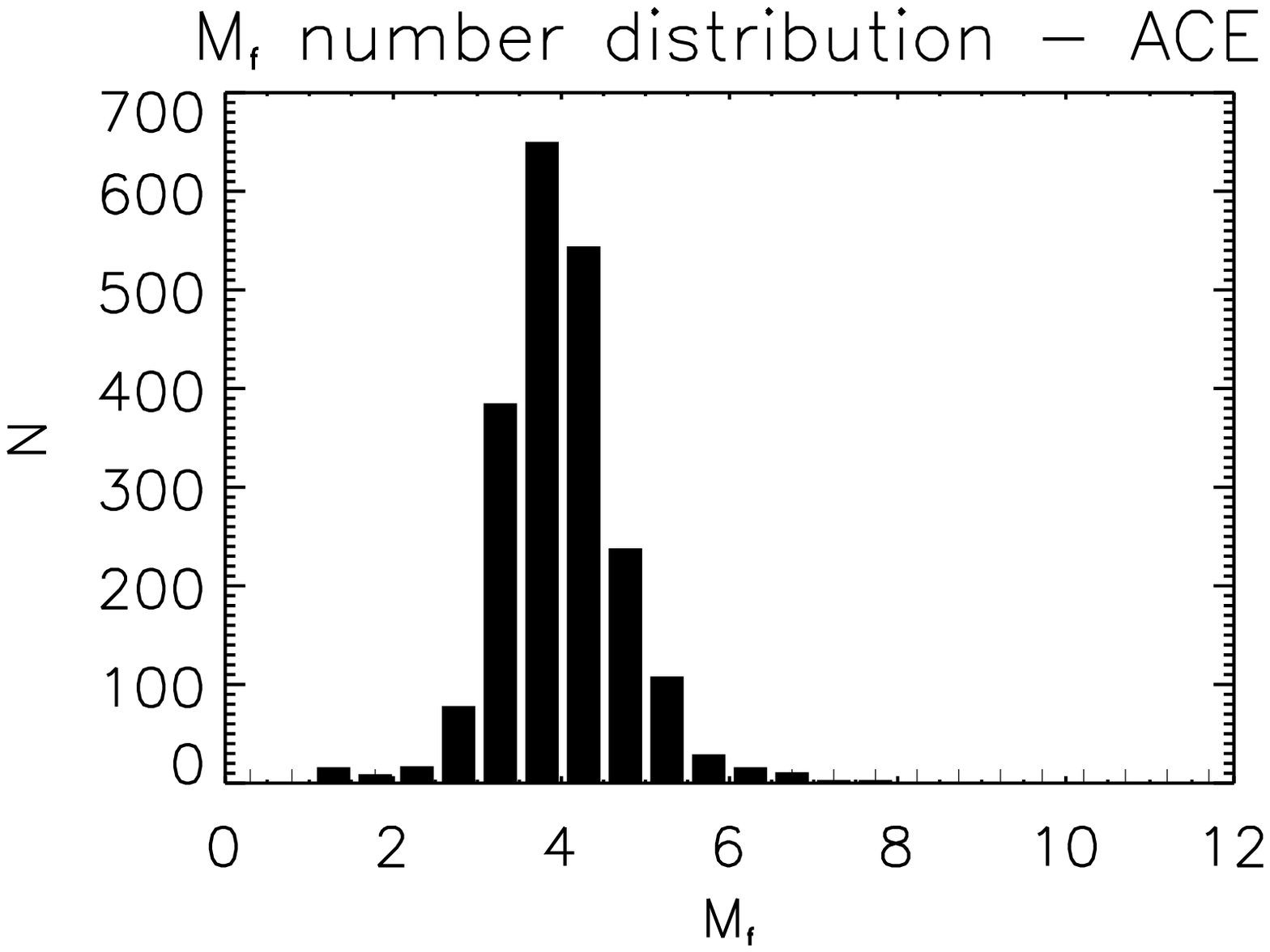, width=150pt} \\
\epsfig{file=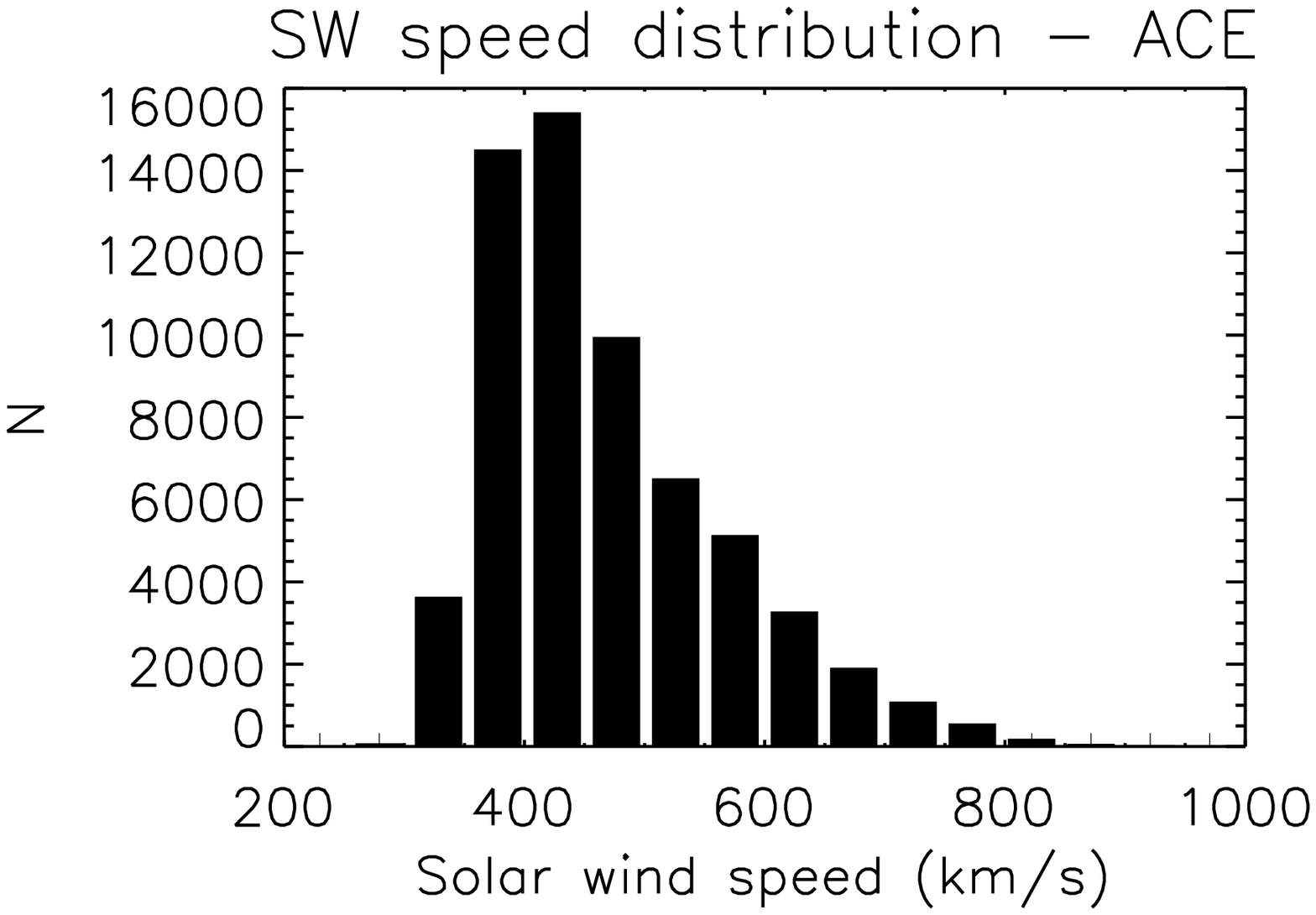, width=150pt} &
\epsfig{file=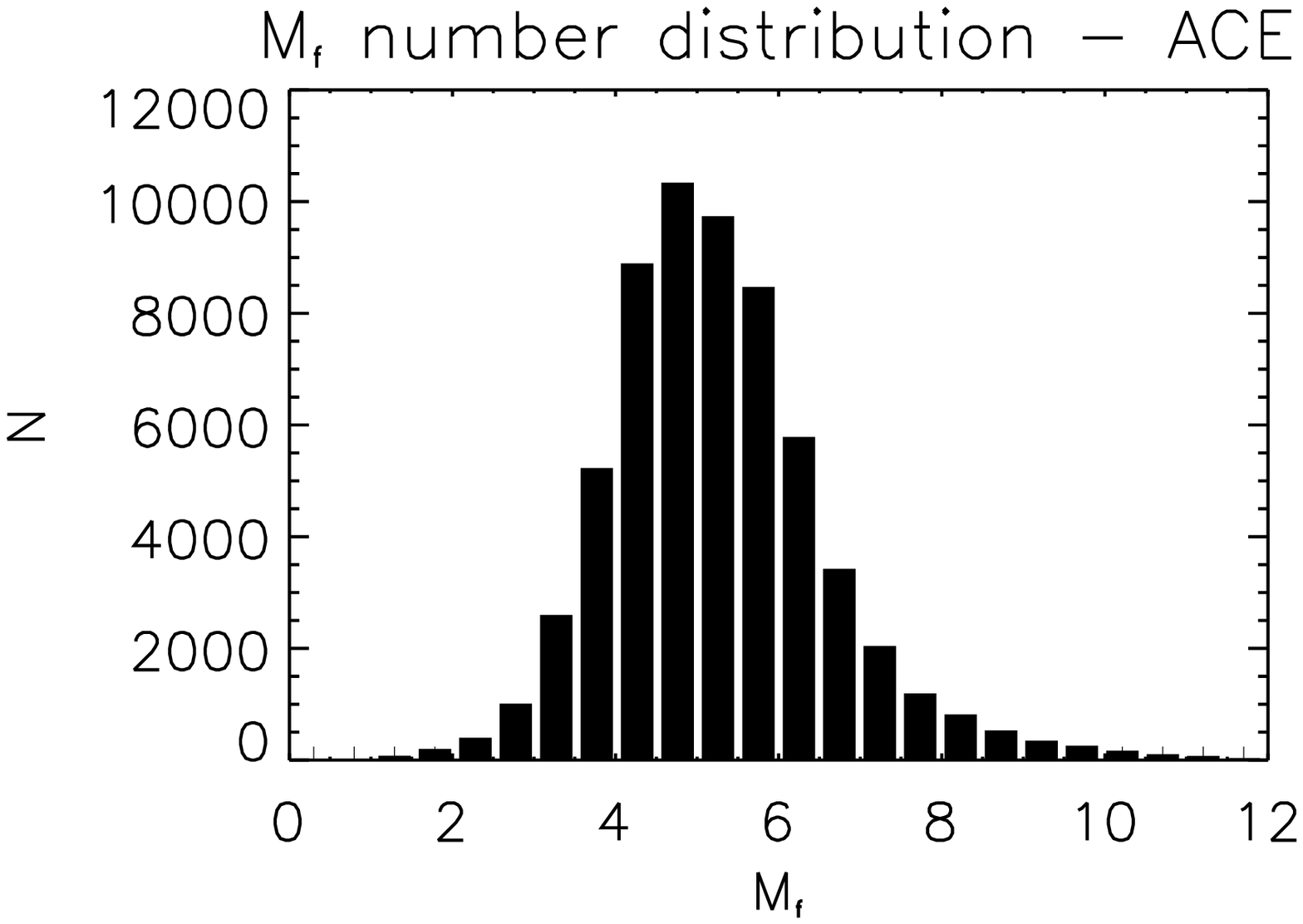, width=150pt} \\
\end{tabular}
\caption{\textit{Top left: SW speed magnitude distribution before HFAs based on the measurement of Cluster-1 CIS; top right: same for Cluster-3. Second panel left: ACE solar wind speed (SWEPAM) for all HFA intervals, right: ACE fast magnetosonic Mach numbers ($M_{f}$) for the same intervals. Third, bottom left: the SW speed distribution for the full 3-month and 1997-2006, respectively (ACE), right: in $M_{f}$ Mach number for the same time intervals.}}
\label{fig:sw}
\end{figure}

When estimating the size of HFAs we needed the solar wind speed value (See: \citet{facsko05:_ident_statis_analy_hot_flow}) but only at one point in the undisturbed field before the event. After we recognized the higher speed, we examined the speed variation in longer time intervals. We searched longer (5-30\,minutes) periods without any perturbation in the solar wind and in the magnetic field using ACE SWEPAM and MAG measurements (\textit{Fig.~\ref{fig:sw}, second row, left side}), furthermore, before the bow-shock crossings using Cluster-1 and -3 CIS and FGM data (\textit{Fig.~\ref{fig:sw}, top panels}). The average value of the solar wind speed magnitude before HFAs was $680 \pm 86$\,km/s (Cluster-1), $671 \pm 92$\,km/s (Cluster-3), and $665 \pm 84$\,km/s (ACE), respectively, where all HFAs where included. In contrast, the average solar wind speed during the whole period was much lower, $546 \pm 97$\,km/s (ACE). The latter is still higher than the long-term average due to extended fast wind intervals (\textit{Fig.~\ref{fig:sw}, third row, left side}). Averaging over all measurements of ACE in 1997-2006 (See: \textit{Fig.~\ref{fig:sw}, bottom, left side}) yields a lower average than the one year period's result: $492 \pm 102$\,km/s. This fact also underlines the importance of fast solar wind in the formation of HFAs. 

\begin{figure}[th]
\centering
\epsfig{file=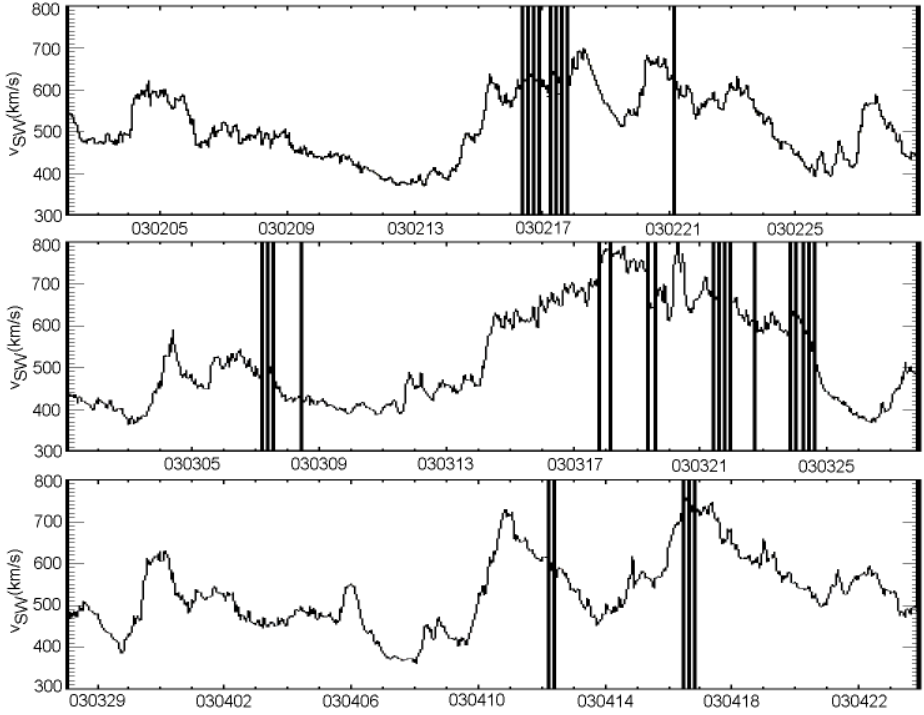, width=350pt}
\caption{\textit{The measured SW speed at the L1 point for 3 solar rotations. ACE 1-hr averages are from ACE Science Center. The times of the observed HFAs are marked by vertical lines.}}
\label{fig:swt}
\end{figure}

By marking the dates of HFA events on a solar wind speed vs. time plot one can recognize that nearly all HFA events appear when the solar wind speed is higher than average (See \textit{Fig.~\ref{fig:swt} marked by vertical lines}). Furthermore, most of events occurred when the solar wind began slowing down within the same co-rotating regions. Since the solar wind density also decreased in these time intervals, the events were coupled to rarefaction regions. This view is supported by the very high fast magnetosonic Mach numbers (\textit{Fig.~\ref{fig:sw}, second row, right side}). If one compares the speed distribution of the HFAs in Mach units to the common solar speed values, its distribution seems to be strange because usually only a small portion of the values is so high. The high value of the mean fast magnetosonic Mach number suggests that the bow shock was in super critical state when these events were observed. Actually, similar phenomena were discovered at Mars \citep{oeieroset01:_hot_martian}. The observation of this new condition and the fact that the fast magnetosonic speed is lower in the outer part of the Solar System suggest that HFAs should be more frequent phenomena near the giant planets and probably they are observable by the instruments of Cassini.

\section{Summary and discussion}
\label{sec:summary}

We confirmed the features of HFAs based on earlier observation and discovered approximately 50 events in the Cluster FGM, CIS and RAPID measurements. The parameters of these events were used to check the simulation results by \citet{lin02:_global}. Although the error of our first size estimation was too high, another one qualitatively confirmed the simulations. The geometrical condition of tangential discontinuities of forming HFAs was also confirmed and a huge gap around the Sun-Earth direction was found in accordance with \citet{schwartz00:_condit} based on ISEE-1, -2, Interball, WIND and IMP-8 measurements. 

The role of fast solar wind seems to be a very important condition of HFA formation. The observed HFA series were nearly all coupled to fast solar wind intervals or more exactly to their rarefaction regions. \citet{safrankova00:_magnet} have already discovered a very similar condition but their solar wind speed values were lower and they identified the HFAs in the magnetosheath. We are not sure totally in the physical reason of this condition however it is well known that the quasi-perpendicular bow-shock turns the direction of a part of solar wind ions with almost $180^o$. This ion population interacts to ions of the unperturbed solar wind and heats the foreshock region. This mechanism accelerates the trapped ions between the TD and the bow-shock. If the solar wind speed magnitude is bellow a limit no HFA forms -- or only a small one is created which cannot be observed so easy. Our opinion is that both Mach number and solar wind speed are important because the acceleration mechanism across the bow-shock is more efficient at higher Mach number and more kinetic energy heats the cavity if the solar wind velocity is higher.

As we pointed out before \citep{kecskemety06:_distr_rapid_clust}, the development of these events is a relatively common process as they seem to form provided that the conditions are fulfilled. The reason of the low number of events detected seems to be an observation problem. The fast magnetosonic speed is lower in the outer solar system -- otherwise the Mach number is higher there which plays role in the particle acceleration mechanism described above -- so we suppose that HFA events occur more often in that region. We suggest performing an investigation for HFAs at Saturn using the Cassini's MAG and CAPS instruments and a similar analysis at Venus using Venus Express MAG and ASPERA measurements. The examination of several years of Cluster data is needed to find out any solar cycle dependence in the occurrence of HFAs

\section*{Acknowledgements}
\label{sec:acknow}

The authors thank the Cluster FGM, CIS and ACE MAG, SWEPAM teams for providing data for this study. This work was supported by OTKA grant T037844 of the Hungarian Scientific Research Fund.

\bibliography{gabor_facsko}
\bibliographystyle{elsart-harv}

\end{document}